\documentclass[aps,prl,preprint,tightenlines,superscriptaddress,showpacs,byrevtex]{revtex4} 
\usepackage{graphicx} 
\usepackage{epsfig}  
\usepackage{amsmath} 
\usepackage[dvips]{color} 
\usepackage[dvips]{pstcol} 
\usepackage{dcolumn}  
\graphicspath{{ps}}

\newcommand{\dppp}{D^0 \to \pi^+\pi^-\pi^0} 
\newcommand{\dkpp}{D^0 \to K^-\pi^+\pi^0} 
\newcommand{\bratio}
{\mathcal{B}(D^0 \to \pi^+\pi^-\pi^0)/\mathcal{B}(D^0 \to K^- \pi^+ \pi^0)} 
\newcommand{\dbar}{\overline{D}{}^0} 

\begin{document} 

\preprint{\vbox{ \hbox{Belle Preprint 2007-49} 
                 \hbox{KEK   Preprint 2007-67} 
}}

\title{ \quad\\[0.5cm]  Measurement of the ratio 
$\mathcal{B}$($D^0 \to \pi^+\pi^-\pi^0$)/
$\mathcal{B}$($D^0 \to K^-\pi^+\pi^0$) and 
the time-integrated $CP$ asymmetry in $D^0 \to \pi^+\pi^-\pi^0$ } 

\vspace{1cm} 

\begin{abstract} 

We report a high-statistics measurement of the relative branching fraction 
$\bratio$ using a 532 fb$^{-1}$ data sample collected with the Belle detector at 
the KEKB asymmetric-energy $e^+ e^-$ collider. 
The measured value of the relative branching fraction is 
$\bratio \ = \ (10.12 \pm 0.04({\rm stat}) \pm 0.18({\rm syst})) \times 10^{-2}$, 
which has an accuracy comparable to the world average. 
We also present a measurement of the time--integrated $CP$ 
asymmetry in $D^0 \to \pi^+\pi^-\pi^0$ decay. 
The result, $A_{CP} \ = \ (0.43 \ \pm \ 1.30)\%$, shows no significant $CP$ violation. 
\end{abstract} 


\affiliation{Budker Institute of Nuclear Physics, Novosibirsk}
\affiliation{Chiba University, Chiba}
\affiliation{University of Cincinnati, Cincinnati, Ohio 45221}
\affiliation{Justus-Liebig-Universit\"at Gie\ss{}en, Gie\ss{}en}
\affiliation{The Graduate University for Advanced Studies, Hayama}
\affiliation{Hanyang University, Seoul}
\affiliation{University of Hawaii, Honolulu, Hawaii 96822}
\affiliation{High Energy Accelerator Research Organization (KEK), Tsukuba}
\affiliation{Institute of High Energy Physics, Chinese Academy of Sciences, Beijing}
\affiliation{Institute of High Energy Physics, Vienna}
\affiliation{Institute of High Energy Physics, Protvino}
\affiliation{Institute for Theoretical and Experimental Physics, Moscow}
\affiliation{J. Stefan Institute, Ljubljana}
\affiliation{Kanagawa University, Yokohama}
\affiliation{Korea University, Seoul}
\affiliation{Kyungpook National University, Taegu}
\affiliation{\'Ecole Polytechnique F\'ed\'erale de Lausanne (EPFL), Lausanne}
\affiliation{University of Ljubljana, Ljubljana}
\affiliation{University of Maribor, Maribor}
\affiliation{University of Melbourne, School of Physics, Victoria 3010}
\affiliation{Nagoya University, Nagoya}
\affiliation{Nara Women's University, Nara}
\affiliation{National Central University, Chung-li} 
\affiliation{National United University, Miao Li}
\affiliation{Department of Physics, National Taiwan University, Taipei}
\affiliation{H. Niewodniczanski Institute of Nuclear Physics, Krakow}
\affiliation{Nippon Dental University, Niigata}
\affiliation{Niigata University, Niigata}
\affiliation{University of Nova Gorica, Nova Gorica}
\affiliation{Osaka City University, Osaka}
\affiliation{Osaka University, Osaka}
\affiliation{Panjab University, Chandigarh}
\affiliation{Saga University, Saga}
\affiliation{University of Science and Technology of China, Hefei}
\affiliation{Seoul National University, Seoul}
\affiliation{Sungkyunkwan University, Suwon}
\affiliation{University of Sydney, Sydney, New South Wales}
\affiliation{Tata Institute of Fundamental Research, Mumbai}
\affiliation{Toho University, Funabashi}
\affiliation{Tohoku Gakuin University, Tagajo}
\affiliation{Tohoku University, Sendai}
\affiliation{Department of Physics, University of Tokyo, Tokyo}
\affiliation{Tokyo Institute of Technology, Tokyo}
\affiliation{Tokyo Metropolitan University, Tokyo}
\affiliation{Tokyo University of Agriculture and Technology, Tokyo}
\affiliation{Virginia Polytechnic Institute and State University, Blacksburg, Virginia 24061}
\affiliation{Yonsei University, Seoul}
  \author{K.~Arinstein}\affiliation{Budker Institute of Nuclear Physics, Novosibirsk} 
  \author{I.~Adachi}\affiliation{High Energy Accelerator Research Organization (KEK), Tsukuba} 
  \author{H.~Aihara}\affiliation{Department of Physics, University of Tokyo, Tokyo} 
  \author{V.~Aulchenko}\affiliation{Budker Institute of Nuclear Physics, Novosibirsk} 
  \author{T.~Aushev}\affiliation{\'Ecole Polytechnique F\'ed\'erale de Lausanne (EPFL), Lausanne}\affiliation{Institute for Theoretical and Experimental Physics, Moscow} 
  \author{T.~Aziz}\affiliation{Tata Institute of Fundamental Research, Mumbai} 
  \author{S.~Bahinipati}\affiliation{University of Cincinnati, Cincinnati, Ohio 45221} 
  \author{A.~M.~Bakich}\affiliation{University of Sydney, Sydney, New South Wales} 
  \author{V.~Balagura}\affiliation{Institute for Theoretical and Experimental Physics, Moscow} 
  \author{E.~Barberio}\affiliation{University of Melbourne, School of Physics, Victoria 3010} 
% ...
\collaboration{The Belle Collaboration}

\maketitle 

\tighten 

{\renewcommand{\thefootnote}{\fnsymbol{footnote}}}
\setcounter{footnote}{0} 

\section{Introduction} 

Using a large data sample of $D^0$ decays 
accumulated with the Belle detector, we obtain a precise determination 
of the $\dppp$ branching fraction using the 
$D^0 \to K^-\pi^+\pi^0$ decay mode for normalization~\cite{kplus}. 
This study 
is the first step towards a high-statistics 
Dalitz-plot analysis of the $D^0 \to \pi^+\pi^-\pi^0$ decay. 
The latter could give insight into the controversy concerning 
the S-wave $\pi^+\pi^-$ contribution in $D$ meson decays~\cite{e791,focus,mura,cleo2} 
and provide a sensitive measurement of $CP$ violation in the neutral $D$ meson 
system. Knowledge of $\mathcal{B}$($D^0 \to \rho \pi$)/$\mathcal{B}$($D^0 \to K^* K$), 
also based on the $D^0 \to \pi^+\pi^-\pi^0$ Dalitz analysis, could improve our 
understanding of the apparent discrepancy of the measured two-body branching 
fractions for $D^0 \to \ K\overline{K}$ and $\pi\pi$ with theoretical 
expectations~\cite{bigi}. A detailed study of the 
$D^0 \to \pi^+\pi^-\pi^0$ decay as well as of other $D^0$ $CP$-symmetric 
final states, can be used to further improve statistics for the measurement 
of the angle $\phi_3$ of the CKM matrix. \\ 

Since both $\dppp$ and $\dkpp$ involve a neutral pion and the same number of 
charged tracks in the final state, several sources of systematic uncertainties 
nearly cancel 
in the determination of the relative branching fraction. 
The method used minimizes any dependence on the assumed decay model. 
The result obtained is compared to recent measurements by the 
CLEO~\cite{cleo3} and BaBar~\cite{babar} collaborations. \\ 

In this study, we also subdivide the same data into $D^0$ and $\dbar$ subsamples 
to calculate the time--integrated $CP$--asymmetry ($A_{CP}$) in the $\dppp$ 
decay mode. 
The latter study is motivated by the recent measurements of mixing 
parameters in neutral $D$--meson system~\cite{mixing}. 
The rate of $CP$ violation predicted by the Standard Model
reaches $\sim 0.1 \%$ in some Cabibbo-suppressed decays such as
$D^0 \to \pi^+ \pi^-\pi^0$~\cite{bigi,nir}. 
The value of $A_{CP}$ in the $D^0 \to \pi^+ \pi^-\pi^0$ decay obtained in the single 
existing measurement by the CLEO collaboration is $(1 ^{+10}_{-9})\%$~\cite{cleo2}. 
We provide a significantly improved measurement 
of $A_{CP}$($D^0 \to \pi^+\pi^-\pi^0$). 
This measurement is complementary to other measurements of $A_{CP}$ in 
singly--Cabibbo suppressed decay modes (the most sensitive is one by the 
BaBar experiment in $D^0 \to K^+K^-,~\pi^+\pi^-$~\cite{babar_kkpp}). \\ 

\section{Experiment} 

The Belle detector is a large-solid-angle magnetic spectrometer 
located at the KEKB $e^+ e^-$ storage rings, which collide 8.0 GeV 
electrons with 3.5 GeV positrons to produce 
$\Upsilon$(4S) at the energy of 10.58 GeV~\cite{kekb}. Closest to 
the interaction point is a silicon vertex detector (SVD) 
surrounded by a 50-layer central drift 
chamber (CDC), an array of aerogel Cherenkov counters (ACC), 
a barrel-like arrangement of 
time-of-flight (TOF) scintillation counters, and an electromagnetic 
calorimeter (ECL) comprised of CsI (Tl) crystals. These subdetectors 
are located inside a superconducting solenoid coil 
that provides a 1.5 T magnetic field. An iron flux-return yoke located 
outside the coil is 
instrumented to detect $K^0_L$ mesons and identify muons. 
The detector is described in detail elsewhere~\cite{det,svd}. \\ 

\section{Data selection}

For this analysis, we used a data sample of 532 fb$^{-1}$. 
To reduce backgrounds and also tag the flavor of the $D^0$ or $\dbar$ 
decay, we require that the $D^0$'s originate from $D^* \to D^0 \pi$ decays. 
A $D^{*\pm}$ candidate
is reconstructed from a $D^0$ and a low momentum $\pi$ where the charge of the 
latter tags the $D^0$ flavor: ${D^*}^+ \ \to \ D^0 \pi^+_{\rm tag}$, 
${D^*}^- \ \to \ \dbar \pi^-_{\rm tag}$~\cite{dstar}. 
$D^0$ candidates are reconstructed from combinations of two oppositely charged 
pions (a pion and a kaon in the case of $\dkpp$) and one neutral 
pion formed by two photons. In the case of multiple candidates, 
we choose the best candidate using a $\chi^2$ value based on the 
vertex information of all charged particles, $M(D^*)-M(D^0)$, and $M(\pi^0)$ values. 
A fit in which the $\pi^{\pm}/K^{\pm}$, $\pi^0$ momenta are constrained to originate 
from a common vertex and have the nominal mass of the $D^0$ meson is also performed. \\ 

The following kinematic and topological criteria are applied to the charged 
track candidates: the distance from the nominal interaction point to the point 
of closest approach of the track is required to be within 0.2 cm in the radial 
direction 
and 2.0 cm along the beam direction. 
We also require the transverse momentum of the track 
to be greater than $0.050\,\mathrm{GeV}/c$, 
to suppress beam background. Kaons and pions are separated by combining the 
responses of the ACC and the TOF with the $dE/dx$ measurement from the CDC to
form a likelihood $\mathcal{L}(h)$, where $h$ is a pion or a kaon.
Charged particles are identified as pions or kaons using the likelihood ratio
$\mathcal{R}=\mathcal{L}(K)/(\mathcal{L}(K)+\mathcal{L}(\pi))$.
For the identification of a charged pion, we require $\mathcal{R}<0.4$; 
this requirement selects pions with an efficiency of 93\% and a kaon 
misidentification probability of 9\%. For the identification of charged kaons, 
the requirement is $\mathcal{R}>0.6$; in this case, the efficiency 
for kaon identification is 86\% and the probability to misidentify a pion is 4\%. 
We require $\theta_{\rm lab}(\pi^{\pm}/K^{\pm}) \ < \ 2.2$ rad to improve 
$K/\pi$ separation~\cite{kpi}, where $\theta_{\rm lab}$ is the angle 
between the particle momentum and the z-axis, defined as the direction 
opposite to that of the positron beam. 
To suppress random combinations of two photons, 
we impose conditions on the energies of the photons constituting the $\pi^0$ 
candidate ($E_{\gamma}(\pi^0)$ $>$ 0.070~GeV), the two-photon invariant mass 
($0.120\,\mathrm{GeV}/c^2 < M(\gamma\gamma) < 0.150\,\mathrm{GeV}/c^2$, 
which corresponds to 2.8 standard deviations ($\sigma$) in reconstructed 
$M(\gamma\gamma)$) and the $\pi^0$ momentum in the laboratory frame 
($p_{\rm lab}(\pi^0) > 0.35\,\mathrm{GeV}/c$) to suppress random combinations 
of two photons. The mass difference of $D^*$ and $D^0$ candidates must satisfy 
the restriction $ 0.1449\,\mathrm{GeV}/c^2 < M(\pi_{\rm tag}\pi^+\pi^-\pi^0) - 
M(\pi^+\pi^-\pi^0) < 0.1461\,\mathrm{GeV}/c^2$ ($2 \sigma$ in 
reconstructed $M(D^*)-M(D^0)$). The momentum of the $D^*$ in the 
center-of-mass (cms) frame of the $\Upsilon(4{\rm S})$ must be in the range 
$ 3.0\,\mathrm{GeV}/c < p_{\rm cms}(D^*) < 4.5\,\mathrm{GeV}/c$. 
The lower cut is applied to reject $D^*$'s originating from $B$ mesons. 
The upper cut excludes the region of $p_{\rm cms}({ D}^*)$ with the largest 
discrepancy between Monte Carlo (MC) simulation and data 
(the difference is taken into account in the systematic error). 
To eliminate background from $D^0 \to K_{\rm S}\pi^0 \to (\pi^+\pi^-)\pi^0$ decays, 
the following veto on $M(\pi^+\pi^-)$ is applied: 
$0.455\,\mathrm{GeV}/c^2 < M(\pi^+\pi^-) < 0.537\,\mathrm{GeV}/c^2$ ($6.5 \sigma$ 
in the reconstructed $K_S$ invariant mass resolution). We also require that the 
$\pi^+\pi^-\pi^0$/$K^-\pi^+\pi^0$ invariant mass be in the range 1.79--1.91 GeV/$c^2$, 
which corresponds to $5.5 \sigma$ in the $M(D^0)$ resolution. For events passing this 
requirement, the momenta of the final state particles are refitted using the nominal 
$D^0$ mass as a constraint. These refitted momenta are used to calculate Dalitz plot 
variables as described below. After applying all selection criteria, we find 
$123.2\times 10^3$ $\dppp$ and $1221.0\times 10^3$ $\dkpp$ 
events in our data sample. 

\section{Efficiency calculation} 

To obtain reconstruction efficiencies, $22 \times 10^6$ MC events, uniformly 
distributed over the Dalitz plane (DP), were generated 
for each of the two modes. They were processed using the GEANT detector 
simulation package~\cite{geant} and reconstructed 
with the same selection criteria as for data. To take into account the 
radiative tail in the $D^0$ invariant mass distribution (Fig.~1) due to final 
state radiation (FSR), the PHOTOS package~\cite{photos} was used for $\dppp$/$\dkpp$ at 
the generator level. \\ 

Differences in the efficiency of particle identification 
(PID) selection criteria between MC and data events are taken into 
account as correction weights to each signal event. They are obtained 
using a large sample of $D^* \to D^0 \pi_{\rm tag}, D^0 \to K^- \pi^+$ decays, 
as a function of the momentum and polar angle of the decay products. The 
uncertainties of these corrections contribute to the systematic uncertainty 
of the result. We apply these weights only to the kaon in $\dkpp$ and to the 
corresponding pion (of the same charge) in $\dppp$, 
since the corrections to the PID efficiency for the remaining decay pion 
and the tagging pion ($\pi_{\rm tag}$) cancel in the ratio. \\ 

A certain portion of signal MC events ($\sim 15\%$) is reconstructed with one or more 
random particles ($\gamma$'s, $\pi^0$'s, $\pi^{\pm}$ or $K^{\pm}$) combined with 
true signal daughters. We distinguish correctly 
reconstructed and misreconstructed signal MC events by comparing the reconstructed 
momenta of all the final state particles to the corresponding generator information. 
The correctly reconstructed MC events are used to calculate the 
reconstruction efficiency, and the misreconstructed decays are treated as an 
additional source of background. \\ 

The $M(D^0)$ distributions 
for correctly reconstructed signal MC events 
(Fig.~1a) are fitted with a double hyperbolic Gaussian~\cite{hyper} 
and one regular Gaussian. The $M(D^0)$ distributions for the 
misreconstructed signal MC events (Fig.~1b) are fitted with a triple 
Gaussian. The results of the $M(D^0)$ fits are used to fix the shape of 
the signal and misreconstructed signal events for the data $M(D^0)$ fit. \\ 

\begin{figure}[ht!] 
 \epsfig{figure=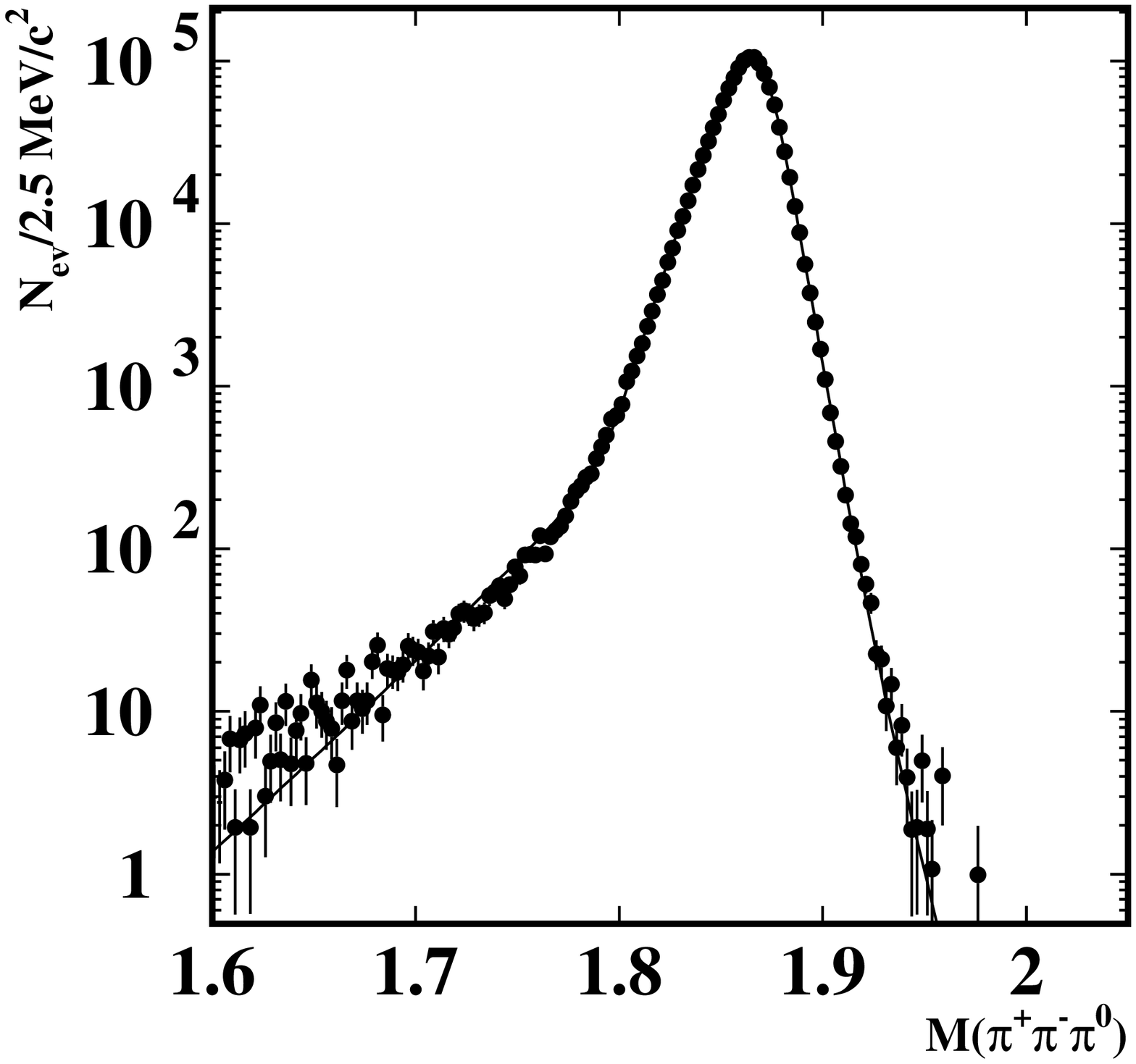,width=0.48\textwidth} 
 \epsfig{figure=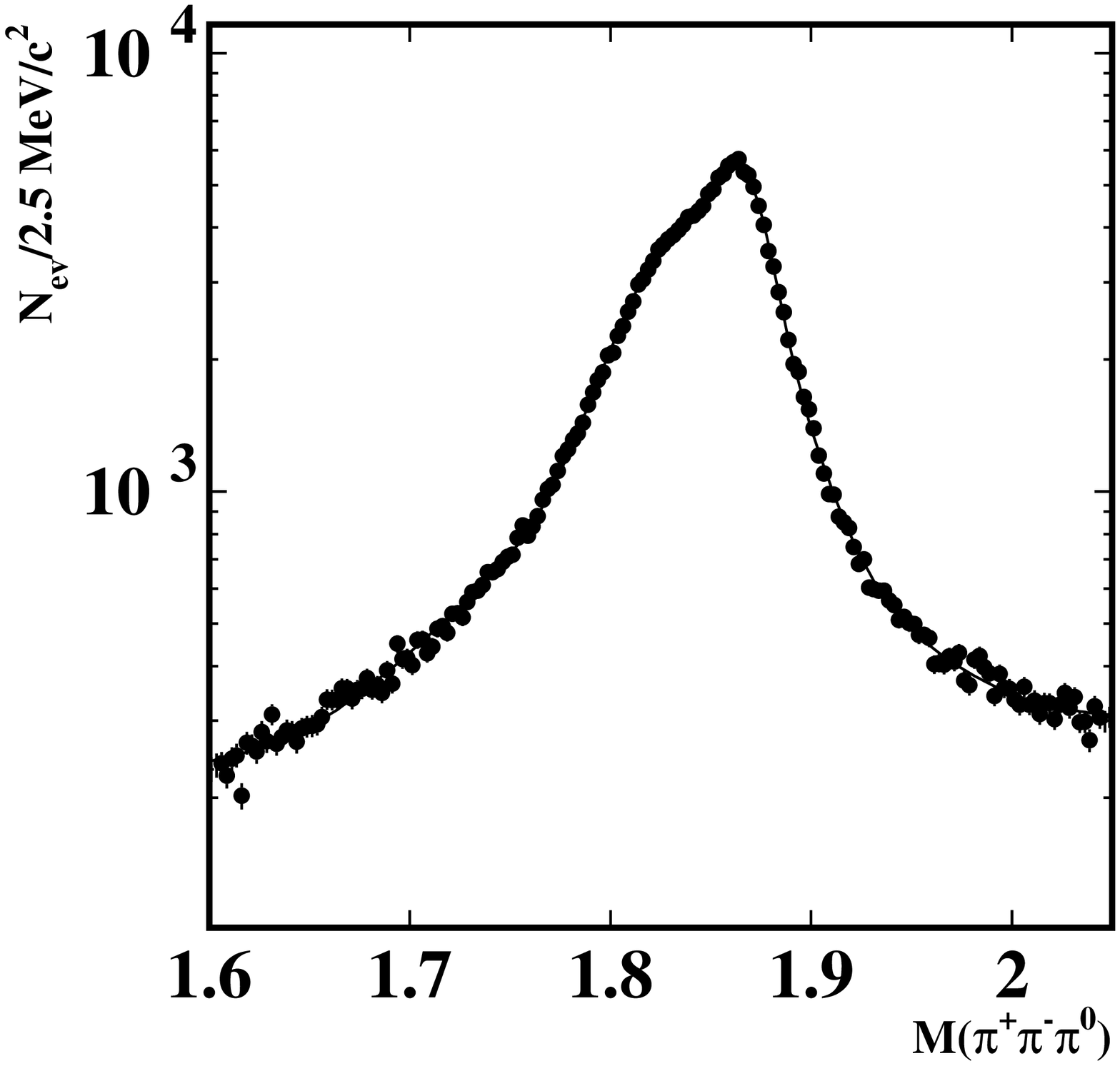,width=0.48\textwidth} 
 \put(-405,190){\bf a)} 
 \put(-175,190){\bf b)} 
 \caption{{\bf (a)} correctly reconstructed ($\chi^2/n.d.f. = 1.7$) and 
 {\bf (b)} misreconstructed ($\chi^2/n.d.f. = 1.0$) signal MC distributions.} 
\end{figure} 

The fraction of correctly reconstructed events in a certain bin depends on its 
position on the DP, i.e. $M^2(h\pi)$ vs. $M^2(\pi\pi^0)$ ($h$ is $K$ or $\pi$). 
To determine the reconstruction efficiency we divide the DP into bins of size 
0.1 GeV$^2$/$c^4$ $\times$ 0.1 GeV$^2$/$c^4$, accumulate signal MC events from 
the $M(D^0)$ signal region, and then normalize by the number of generated events 
in each bin. The calculated values are later used as reciprocal weights for the 
corresponding data distribution. This method takes into account variations of the 
DP data density and minimizes $D^0$ decay model dependence. \\ 

\section{Background study} 

To describe the shape of background in the $M(D^0)$ signal region
for $\dppp$ and $\dkpp$, a sample of generic MC events (including all significant 
processes in $e^+ e^-$ production of $\Upsilon(4S)$, $u\bar{u}$, $d\bar{d}$, 
$s\bar{s}$ and $c\bar{c}$ at the given $\sqrt{s}$), equivalent to 
$\sim$600 fb$^{-1}$, was processed with the same selection criteria as data. 
All generic MC events reconstructed as $\dppp$ were separated into three types: 
contributions from $u\bar{u}$, $d\bar{d}$, and $s\bar{s}$ fragmentation, 
and $\Upsilon(4S) \to B\overline{B}$ events; 
a contribution from $D^* \to D^0(K^-\pi^+\pi^0) \pi_{\rm tag}$ ($c\bar{c}$ events) 
where the charged kaon is misidentified as a pion (the largest source of background); 
and a contribution from $c\bar{c}$ background that does not involve particle 
misidentification and from which the signal is excluded (Fig.~2a -- 2c). For the 
$\dkpp$ case, there are contributions from $u\bar{u}$, $d\bar{d}$, $s\bar{s}$ 
and $\Upsilon(4S) \to B\overline{B}$ events, a contribution from 
$D^* \to D^0(K^-\pi^+\pi^0\pi^0) \pi_{\rm tag}$ via $D^0 \to K^* \rho$ 
and $D^0 \to a_1 K$, as well as a small residual $e^+ e^- \to c\bar{c}$ 
background (Fig.~2d -- 2f). The small peak in the signal region of the latter 
(Fig.~2f) is mainly due to combinations of a $D^0$ and a random 
$\pi_{\rm tag}$ and has to be taken into account. This background is also 
present in $\dppp$. As described previously, the contributions of 
misreconstructed signal MC events are treated as separate sources of background 
for both decay modes. 

\begin{figure}[ht!] 
 \epsfig{figure=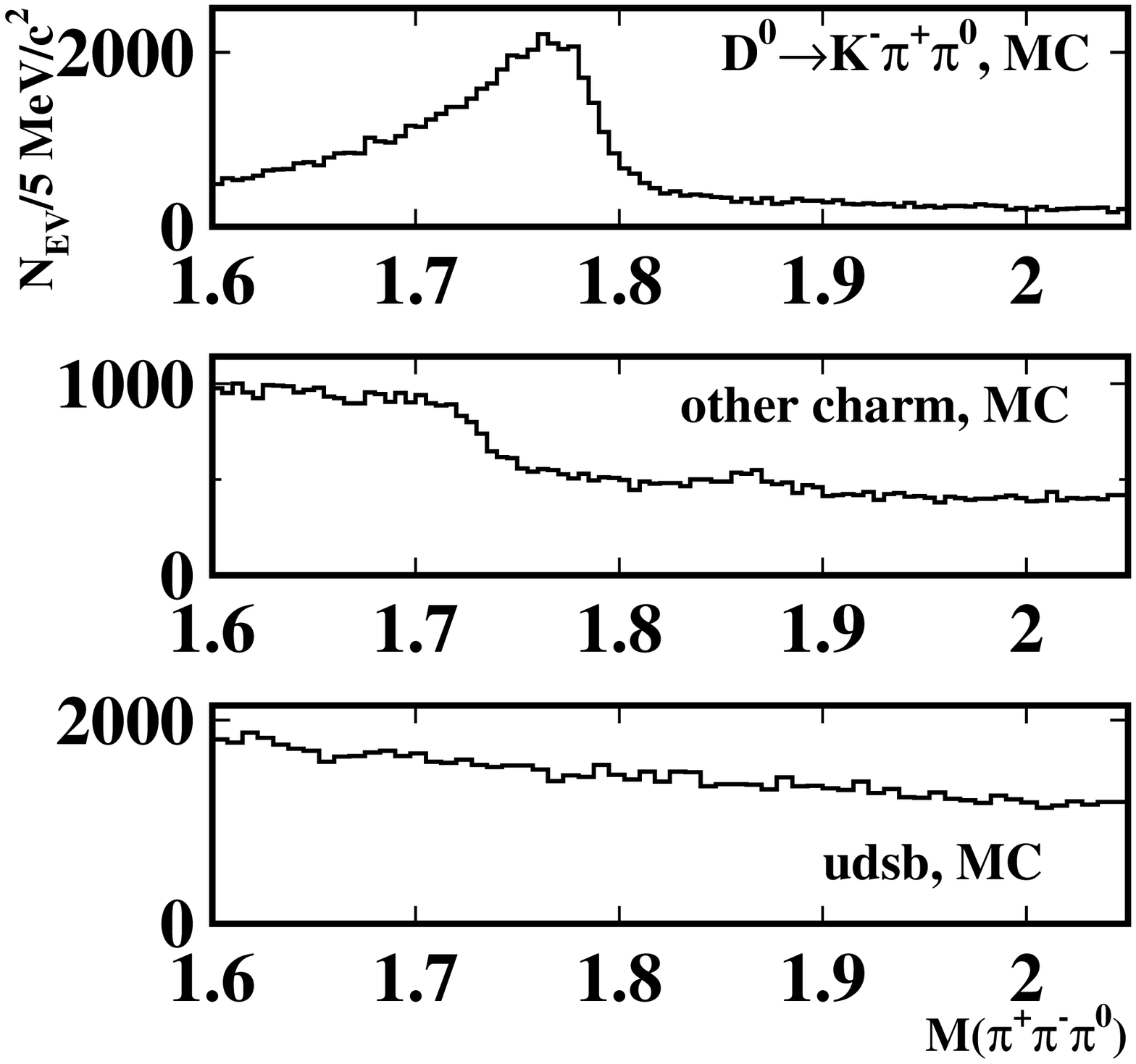,width=0.48\textwidth}  
 \epsfig{figure=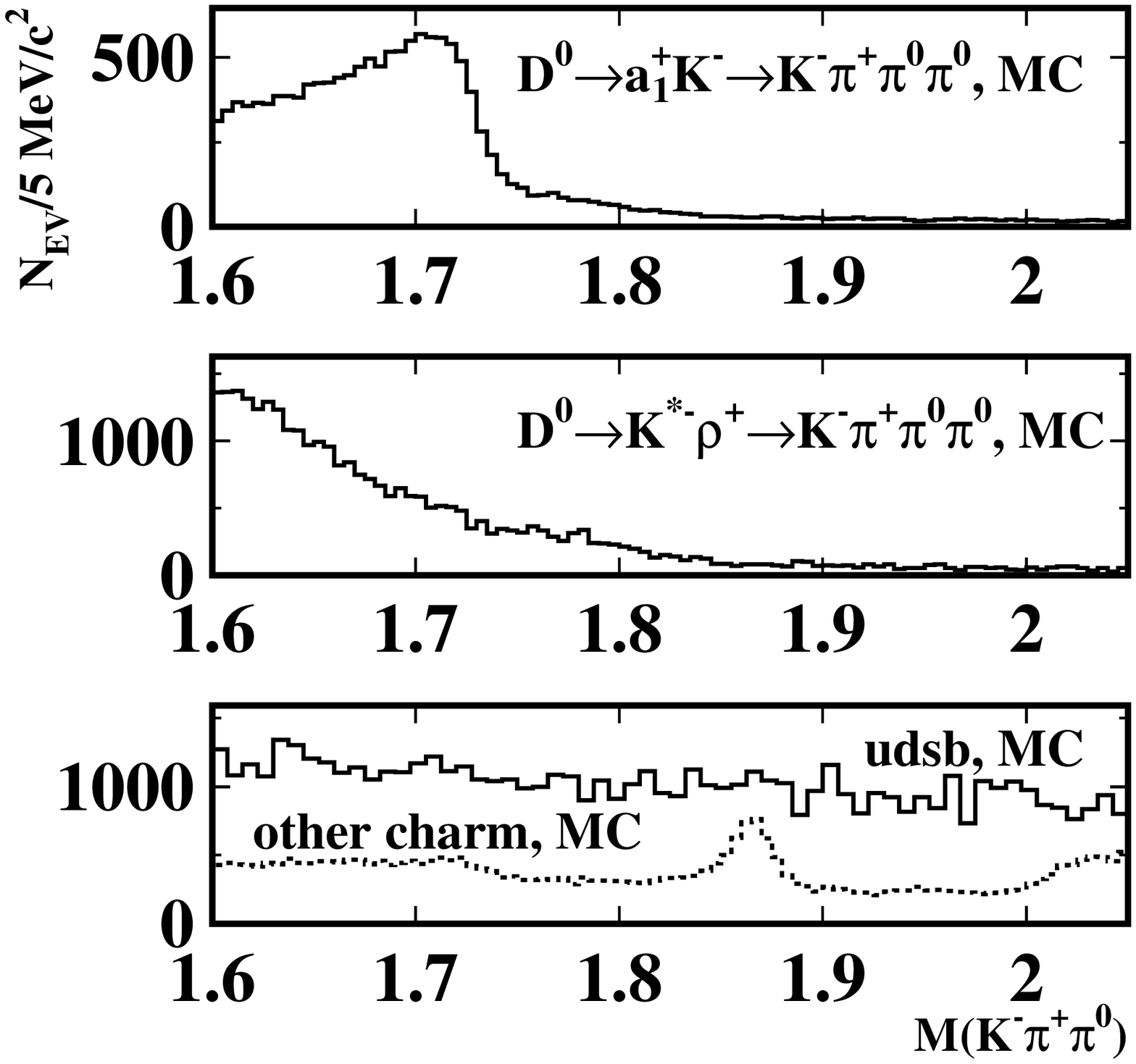,width=0.48\textwidth}  
 \put(-408,200){\bf a)}
 \put(-408,120){\bf b)}
 \put(-408,50){\bf c)}
 \put(-178,185){\bf d)}
 \put(-178,110){\bf e)}
 \put(-178,40){\bf f)}
\caption{
$M(D^0)$ distributions for MC background events in (a-c) the 
$\dppp$ and (d-f) the $\dkpp$ sample: 
{\bf (a)} $D^0 \to K^-\pi^+\pi^0$ events, 
{\bf (b)} other $e^+ e^- \to c\bar{c}$ contributions, 
{\bf (c)} contributions from light quark and $B\overline{B}$ decays. 
{\bf (d)} $D^0 \to a_1 K \to K^-\pi^+\pi^0\pi^0$ events, 
{\bf (e)} $D^0 \to K^* \rho \to K^-\pi^+\pi^0\pi^0$ events, 
{\bf (f)} other $e^+ e^- \to c\bar{c}$ contributions and 
contributions from light quark and $B\overline{B}$ decays. 
Events from the $M(D^0)$ signal region 
(1.79 to 1.91 GeV) are selected for the branching fraction calculation. } 
\end{figure}  
    
\section{Data $M(D^0)$ fit} 

The $M(D^0)$ distribution in data is fitted using fixed MC shapes for 
the various background components, and a signal shape that allows 
for data-MC differences. The fit function for $\dppp$ is 
\begin{equation}
\begin{split}
& F_1 \ = \ N_{\rm sig}\times P_{\rm sig}(\sigma_{\rm add},\Delta x) 
        \ + \  N_{\rm misrec}\times P_{\rm misrec} \ + \ \\ 
& \ + \  N_{udsb}\times P_{udsb} 
  \ + \ N_{\rm misid}\times P_{\rm misid} \ + \  N_{c}\times P_{c}, 
\end{split} 
\end{equation} 

\noindent where $P_{\rm sig}$ and $P_{\rm misrec}$ are the shapes of the $M(D^0)$ 
distributions for correctly reconstructed and misreconstructed signal MC events 
obtained from the corresponding MC distributions. 
The $M(D^0)$ shapes of $u,d,s$--quark and $B\overline{B}$ decays, misidentified 
$D^0 \to K^-\pi^+\pi^0$ decays and other $c$--quark contributions are denoted as 
$P_{udsb}$, $P_{\rm misid}$ and $P_{c}$, respectively; 
$N_{\rm sig}$, $N_{\rm misrec}$, $N_{udsb}$, $N_{\rm misid}$ and 
$N_{c}$ are the normalizations of all the event types and are 
free parameters in the fit. The additional free parameter $\Delta x$ represents 
a common shift in the central value of the Gaussians describing the signal. 
Similarly, $\sigma_{\rm add}$ is a free parameter added in quadrature to all 
the widths of the Gaussian functions (0.3~MeV for $M(\dppp)$ and 0.1~MeV for 
$M(\dkpp)$). The $M(\dkpp)$ fit function has a similar form: 
\begin{equation} 
\begin{split} 
& F_2 \ = \ N_{\rm sig}\times P_{\rm sig}(\sigma_{\rm add},\Delta x) 
        \ + \ N_{\rm misrec}\times P_{\rm misrec} \ + \ \\ 
&   \ + \  N_{udsb}\times P_{udsb}  \ + \ N_{K^*\rho}\times P_{K^*\rho} 
  \ + \  N_{a_1 K}\times P_{a_1 K} \ + \ N_{c}\times P_{c}, 
\end{split} 
\end{equation} 
where $P_{K^*\rho}$ and $P_{a_1 K}$ are the shapes of the contributions of the 
$D^0 \to K^*\rho$ and $D^0 \to a_1 K$ decays to the $M(D^0)$ distribution, respectively, 
and $N_{K^*\rho}$, $N_{a_1 K}$ are their floating normalizations. 
All other variables are the same as in Eq.~(1). 
Figure~3a shows the fit for $M(\dppp)$ described above, while Figure~3b shows that
for $M(\dkpp)$. The low fit quality ($\chi^2/n.d.f. = 3.2$) of the latter is due to 
the large statistics of the signal as well as the 
poor agreement of data and MC simulation for the $D^0 \to K\pi\pi^0\pi^0$ background. 
This discrepancy is taken into account as a systematic error due to the fit uncertainty. \\ 

\begin{figure}[ht!] 
 \epsfig{figure=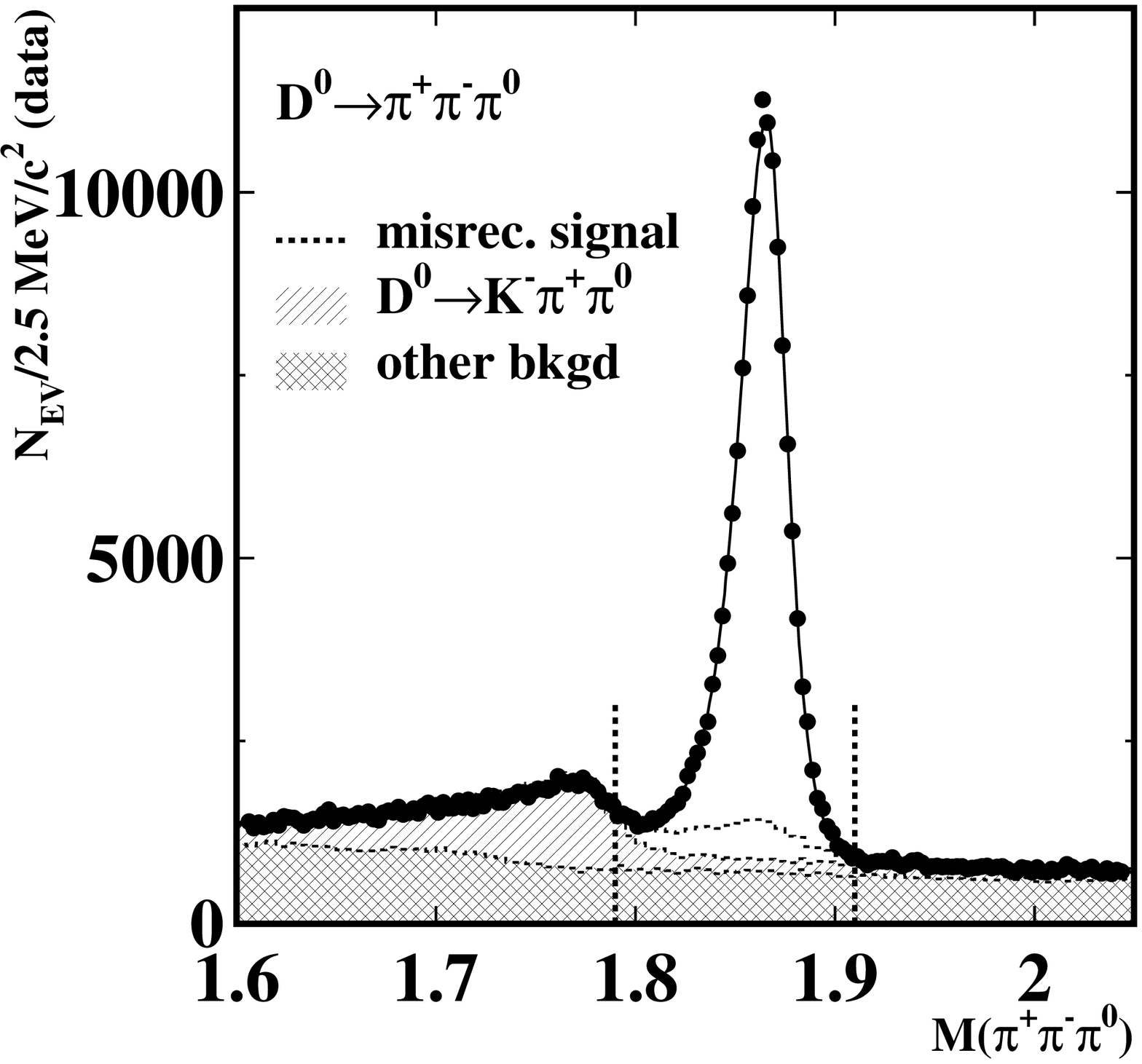,width=0.48\textwidth} 
 \epsfig{figure=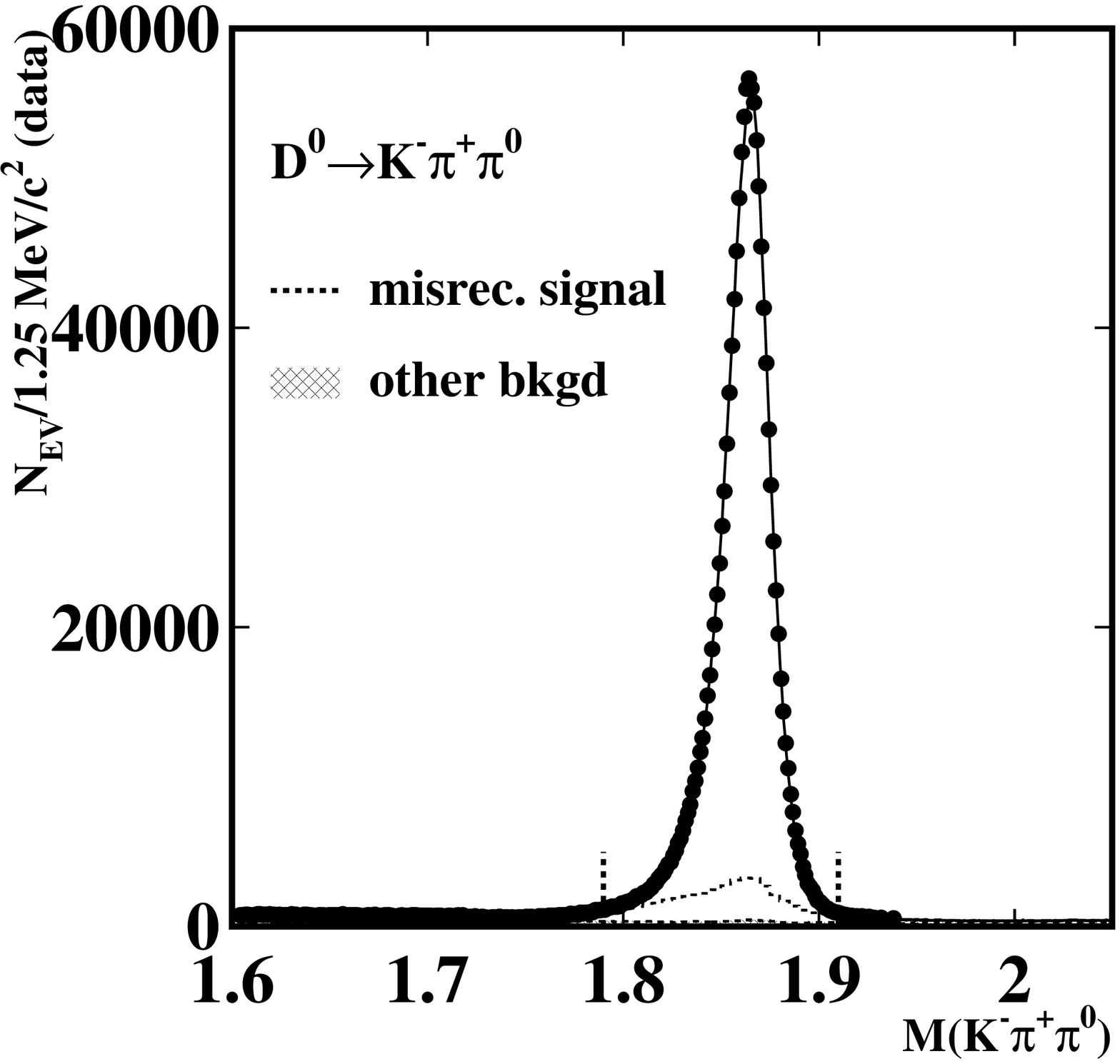,width=0.48\textwidth}
 \put(-260,180){\bf a)}
 \put(-30,180){\bf b)}
   \caption{{\bf (a)} $M(\dppp)$ data fit ($\chi^2/n.d.f. = 1.5$). 
 Data is represented by the points and the curve is the fitted sum of all the 
 contributions (simulated signal and background). The vertical dashed lines 
 indicate the $M(D^0)$ signal region. 
 Background: misreconstructed signal (dashed line), $\dkpp$ with a misidentified 
 kaon (shaded histogram) and other sources, i.e. other $c\bar{c}$ and light quark 
 contributions (hatched histogram). {\bf (b)} $M(\dkpp)$ data 
 fit ($\chi^2/n.d.f. = 3.2$). 
 Background: misreconstructed signal (dashed) and other sources (hatched). 
 The fit results shown correspond to the second step of the $\bratio$ calculation, 
 which takes into account the $D^0$ decay model (see text).} 
\end{figure}

\section{Calculation of the Signal Yields} 

After the parameters of the data $M(D^0)$ distributions are obtained from the fit, 
we fill separate $M^2(h\pi)$ vs. $M^2(\pi\pi^0)$ Dalitz histograms with events from the 
signal $M(D^0)$ region for data and simulated background with the normalizations 
fixed from the fit (Fig.~4). The bin size is 0.1 GeV$^2$/$c^4$ 
$\times$ 0.1 GeV$^2$/$c^4$, the same as for signal MC events. \\ 

The number of $\dppp$ signal events in each bin is calculated as follows:  
\begin{equation} 
\begin{split} 
& Y^i \ = \ D^i  \ - \ N_{\rm misrec}\times S^i_{\rm misrec} \ - \ \\ 
&\ - \ N_{udsb}\times B^i_{udsb} \ - \ N_{\rm misid}\times B^i_{\rm misid} 
  \ - \ N_{c}\times B^i_{c}, 
\end{split} 
\end{equation} 
\noindent 
where $D^i$ is the number of data events, and $N_{\rm misrec}\times S^i_{\rm misrec}$, 
$N_{udsb}\times B^i_{udsb}$, $N_{c}\times B^i_{c}$ 
are the numbers of different background events in the $i$-th bin. 
The procedure for the $\dkpp$ case is similar. \\ 

\begin{figure}[ht!] 
 \epsfig{figure=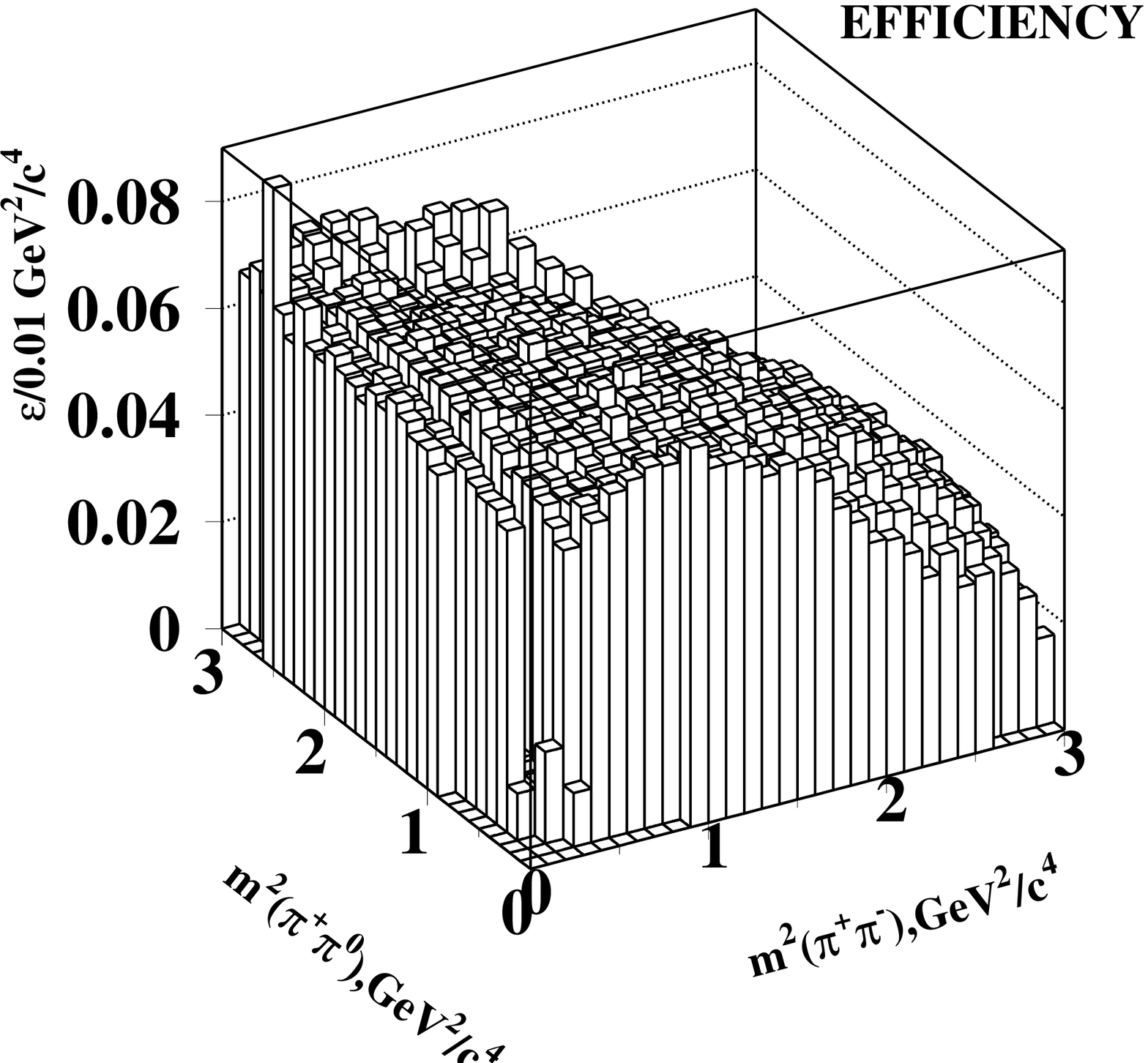,width=0.32\textwidth} 
 \epsfig{figure=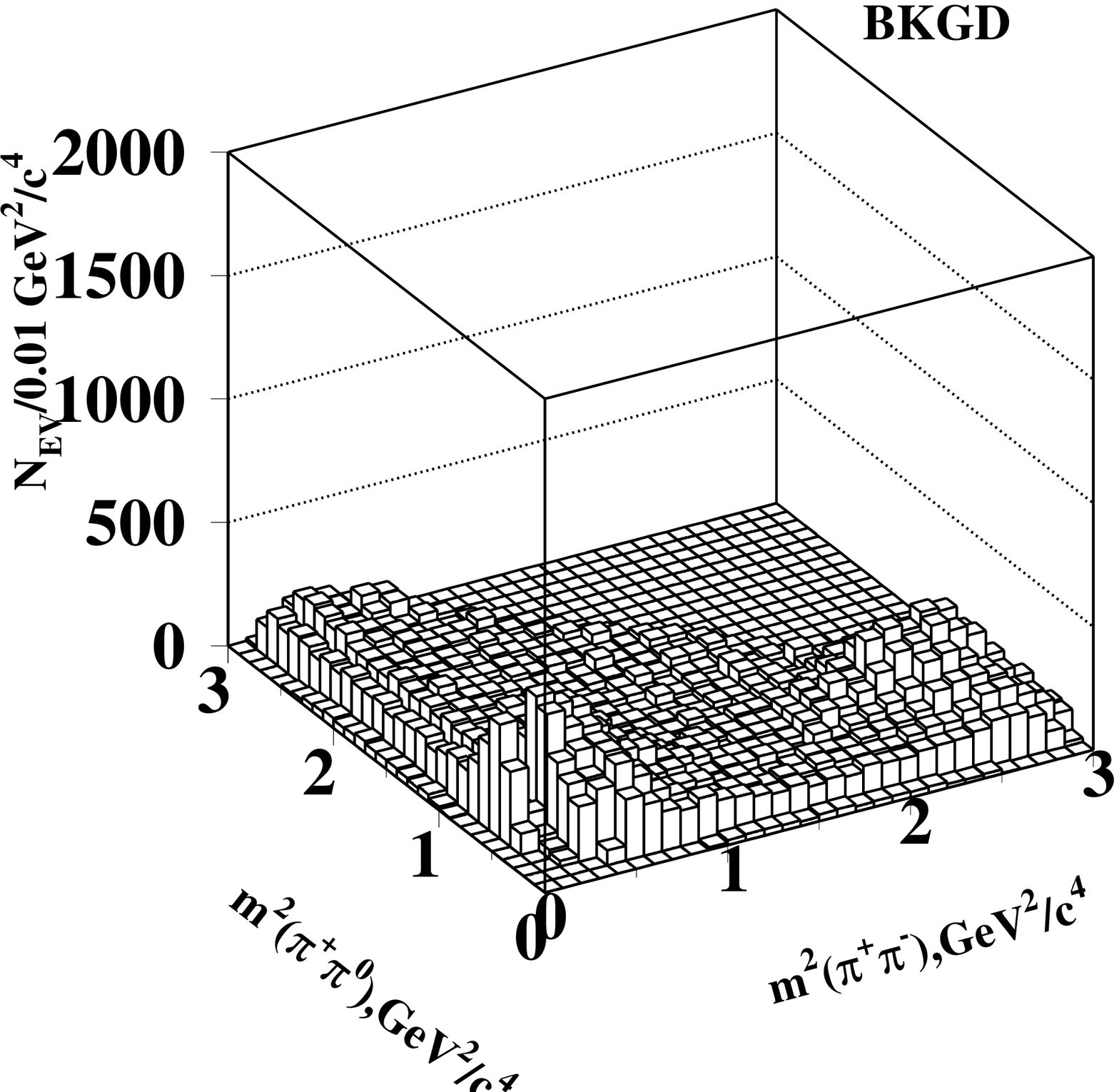,width=0.32\textwidth} 
 \epsfig{figure=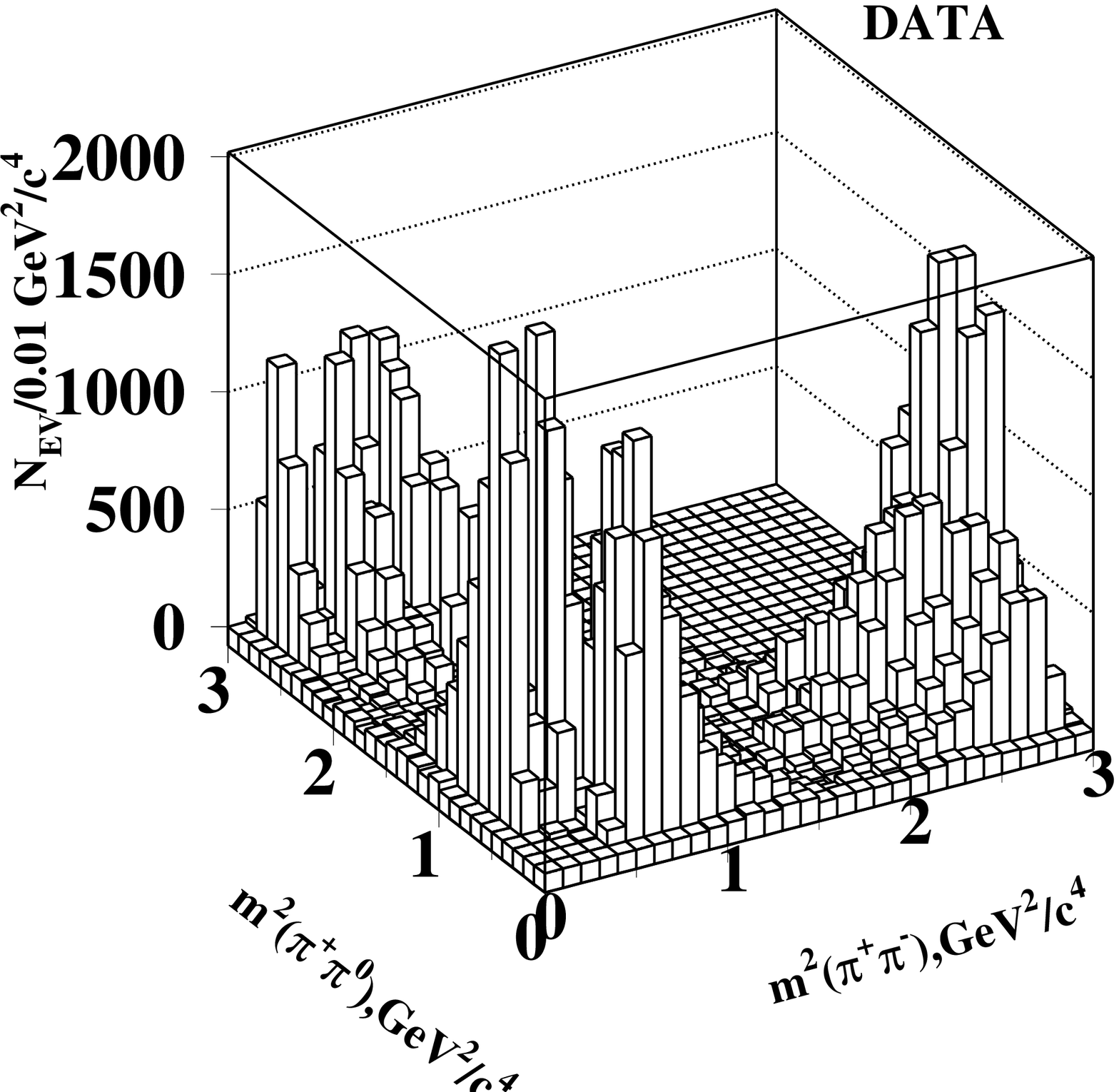,width=0.32\textwidth} 
 \put(-425,145){\bf a)}
 \put(-275,145){\bf b)}
 \put(-120,145){\bf c)} 
 \caption{$\dppp$: Dalitz ($M^2(\pi^+\pi^-)$ vs. $M^2(\pi^+\pi^0)$) 
 distributions for ({\bf a}) efficiency, ({\bf b}) simulated background 
 and ({\bf c}) data.} 
\end{figure} 

The total number of signal events for both decays is
obtained by summing the number $Y^i$ of events over all bins: 
$S \ = \ \sum {Y^i}/{\varepsilon^i},$ where the reconstruction 
efficiency in each bin $\varepsilon^i \ = \ n^i_{\rm rec}/n^i_{\rm gen}$
is used as a reciprocal weight ($n^i_{\rm rec}$ and $n^i_{\rm gen}$ are 
the numbers of reconstructed and generated events in the $i$-th bin). \\ 

At this point, we return to the stage of obtaining the $M(D^0)$ distributions 
from signal MC simulation and perform another iteration of the same procedure using 
the Dalitz histogram for data 
as an approximation of the $D^0$ 
decay model (for each of the two decay modes). As mentioned above, signal MC 
events, used at the first step of our calculations (the entire procedure 
described above), are distributed uniformly over the DP. 
At the second step, the $D^0$ decay model is taken into account by weighting 
entries in a histogram according to their positions on the DP. 
This is done to obtain a more exact $M(D^0)$ distribution for the signal
and misreconstructed signal MC events. 
The distributions are refitted 
and the resulting background normalizations in Eq.~(2) are used to recalculate 
the signal yields $Y^i$ in Eq.~(3). 
This results in 
$S(\dppp) \ = \  (2403.6 \ \pm \ 9.2) \times 10^3$ and   
$S(\dkpp) \ = \  (23751 \ \pm \ 24) \times 10^3$. \\ 

\section{Systematic uncertainties} 

The sources of systematic uncertainty are as follows; 
the values quoted are relative fractions. 
The estimate of the error due to the tracking efficiency uncertainty is based on a large 
sample of partially reconstructed $D^* \to D^0 \pi_{\rm tag}, D^0 \to K_S\pi^+\pi^-$ decays. 
The uncertainty for the two charged tracks --- $\pi^+\pi^-$ or $K^-\pi^+$ --- cancels to 
a large extent in the ratio of the $\dppp$ and $\dkpp$ branching fractions. 
It contributes only 0.01\% to the overall systematic uncertainty.
We assume that the $\pi^0$ and the tagging pion (from the $D^*$) reconstruction 
efficiencies fully cancel in the ratio of the branching fractions. \\ 

The uncertainties of the corrections to the efficiency of PID selection criteria 
contribute $\pm 0.91\%$ to the systematic uncertainty of the result. 
The statistical error of the signal MC sample contributes $\pm 0.30\%$ to the total 
systematic uncertainty. 
The systematic uncertainty due to the fractions of signal and various backgrounds, 
which are fixed from the $M(\dppp)$ fit results, was determined by varying the 
fractions within their errors ($\pm 0.61\%$). 
The correlations between the fit parameters were accounted for 
using the covariance matrix obtained from the fit. 
The uncertainty due to the $M(\dkpp)$ fit ($\pm 0.30\%$) was estimated by relaxing 
or fixing relative normalizations of some of the background types. \\ 

Our method for calculating $\bratio$ minimizes the uncertainty due to 
modelling the $\dppp$ and $\dkpp$ decays. However, the model dependence of the 
background is included in the total systematics. The level of background 
in the $\dkpp$ decay is small and its effect on the ratio of $\bratio$ is 
negligible. The dominant source of background for the $\dppp$ mode 
is the $\dkpp$ decay with the kaon misidentified as a pion. The normalizations 
of the $\dkpp$ submodes ($D^0 \to \rho K$, $D^0 \to K^* \pi$, $D^0 \to {K^*}^0 \pi^0$ 
and non-resonant $\dkpp$) are varied within the uncertainties of their branching 
fractions~\cite{pdg06} and the resulting differences from the central value 
of $\bratio$, summed in quadrature, are treated as the background model 
uncertainty ($\pm 0.48\%$). \\ 

Changing the DP bin size from 0.1 GeV$^2$/$c^4$ $\times$ 0.1 GeV$^2$/$c^4$ 
to 0.05 GeV$^2$/$c^4$ $\times$ 0.05 GeV$^2$/$c^4$ yields a $\pm 0.54\%$ difference in 
the value of $\bratio$. 
We study the effect of the selection criteria upon the fraction of correctly 
reconstructed signal MC events and obtain a corresponding error of $\pm 0.10\%$. 
We varied the event selection criteria in order to estimate the systematic 
error due to any inadequacies in the background description. 
Varying the $K_S$ veto yields a $\pm 0.50\%$ systematic uncertainty.  
The change of $\bratio$ due to the variation of the $p_{\rm cms}$($D^*$) upper cut 
is negligible. Varying the $p_{\rm cms}$($D^*$) 
lower cut 
yields a relatively large uncertainty of $\pm 0.77\%$. 
Uncertainties due to the variation of other selection requirements are listed in 
Table \ref{t1}. The total uncertainty is obtained by adding all contributions 
in quadrature. 

\begin{table}[ht!]  
\caption{Contributions to the relative systematic error on $\bratio$} 
\vspace{0.5cm}  
\label{t1}  
\begin{tabular}{|l|l|l|l|} 
\hline 
Source          & Error, $\%$ & Source             & Error, $\%$ \\ \hline 
PID corrections & 0.91       & {\bf Selection criteria:} &  \\ 
MC statistics   & 0.30       & $K_S$ veto          & 0.50 \\ 
Fit($D^0 \to \pi^+\pi^-\pi^0$)  & 0.61  & $p_{\rm cms}$($D^*$)   & 0.77 \\ 
Fit($D^0 \to K^-\pi^+\pi^0$)    & 0.30  & $M(K^-\pi^+\pi^0/3\pi)$ & 0.36 \\ 
$D^0 \to \pi^+\pi^-\pi^0$ 
backgr. model   & 0.48       & $\Delta M$           & 0.30 \\  
Binning         & 0.54       & $E_{\gamma}$         & 0.40 \\ 
MC misreconstruction            & 0.10 & $M(\pi^0)$ & 0.20 \\ 
Tracking        & 0.01       & $p_{\rm lab}(\pi^0)$ & 0.16 \\ \hline 
\multicolumn{3}{|l|}{\bf Total} & {\bf 1.79} \\ \hline 
\end{tabular} 
\end{table} 

\section{Results for $\bratio$} 

Summarizing the discussion above, we obtain the following ratio of the branching fractions: 
\begin{equation} 
\begin{split} 
& \frac{\mathcal{B}(\dppp)}{\mathcal{B}(\dkpp)} \ = \ \frac{S(\dppp)}{S(\dkpp)} \ = \ \\ 
& \ = \ 0.10120 \pm 0.00040({\rm stat}) \pm 0.00181({\rm syst}) 
  \ = \ 0.1012 \pm 0.0019. 
\end{split} 
\end{equation} 

We can compare our measurement of the ratio with a recent result obtained 
by BaBar~\cite{babar}. There is a $2 \sigma$ difference between the central 
values; the accuracies of the measurements are comparable. 
%
%
To compare results from different experiments, we multiply the obtained value 
of Eq.~(4) by the 2007 world average of 
$\mathcal{B}(D^0 \to K^-\pi^+\pi^0) \ = \ (13.5 \pm 0.6) \%$~\cite{pdg06} 
to calculate the absolute branching fraction for the $D^0 \to \pi^+\pi^-\pi^0$ 
decay (Table \ref{t3}). In a recent study by CLEO~\cite{cleo3}, the relative 
branching fraction $\mathcal{B}(D^0 \to \pi^+\pi^-\pi^0)$/$\mathcal{B}(D^0\to K^-\pi^+)$ 
is measured to be $0.344 \pm 0.005({\rm stat}) \pm 0.012({\rm syst})$. 
Using the world average value of $\mathcal{B}(D^0 
\to K^-\pi^+) \ = \ (3.82 \pm 0.07)\%$ from ~\cite{pdg06}, one can calculate 
the absolute branching fraction of $\mathcal{B}(D^0 \to \pi^+\pi^-\pi^0)$ from CLEO 
data as shown in Table \ref{t3}. A comparison of the corresponding values for
the absolute branching fraction $\mathcal{B}$($D^0 \to \pi^+\pi^-\pi^0$)
shows that the results are in good agreement~\cite{about_newcleo}. 


\begin{table}[ht!] 
\caption{$\mathcal{B}$($D^0 \to \pi^+\pi^-\pi^0$) by Belle, BaBar~\cite{babar} 
and CLEO~\cite{cleo3}. The first two errors are statistical and systematic, 
respectively, and the third one (the fourth column) is the normalization uncertainty. 
The latter is common in the Belle and BaBar results.} 
\vspace{0.5cm} 
\label{t3} 
\begin{tabular}{|l|c|c|c|} 
\hline 
Group              & N$_{\rm ev}, 10^3$ & $\bratio$ 
                                          & $\mathcal{B}(D^0 \to \pi^+\pi^-\pi^0),10^{-3}$ \\ 
                                          \hline 
Belle              & $123.19 \pm 0.49$  & $0.1012 \pm0.0004 \pm 0.0018$ 
                & $13.66 \pm 0.05 \pm 0.24 \pm 0.61 $ \\ 
BaBar              & $60.43 \pm 0.34$   & $0.1059 \pm0.0006 \pm 0.0013$ 
                & $14.30 \pm 0.08 \pm 0.18 \pm 0.64 $ \\ 
CLEO               & $10.83 \pm 0.16$   & --- 
                & $13.14 \pm 0.19 \pm 0.46 \pm 0.24 $ \\ \hline 
\end{tabular} 
\end{table} 

\section{Measurement of $A_{CP}$} 

We subdivide the $\pi^+\pi^-\pi^0$ sample into $\dppp$ and $\dbar\to\pi^+\pi^-\pi^0$ 
subsamples to calculate the value of the time-integrated $CP$-asymmetry using the same 
method for calculating the signal yield that was used for the relative branching 
fraction. The fitted $M(D^0)$ distributions of the data are shown in Fig.~5. 
The resulting values of $S \ = \ \sum {Y^i}/{\varepsilon^i}$ are 
\begin{equation} 
\begin{split} 
& S_{D^0} \ = \ (1154.7 \ \pm \ 6.7)\times 10^3, \\ 
& S_{\dbar} \ = \ (1144.7 \ \pm \ 6.6)\times 10^3. 
\end{split} 
\end{equation} 

Their sum differs from the value used to calculate the branching fraction, because 
the corrections for the PID efficiency of the pions originating from the $D^0$ 
cancel out in the case of the $A_{CP}$ calculation and thus are not applied. \\ 

\begin{figure}[ht!] 
 \epsfig{figure=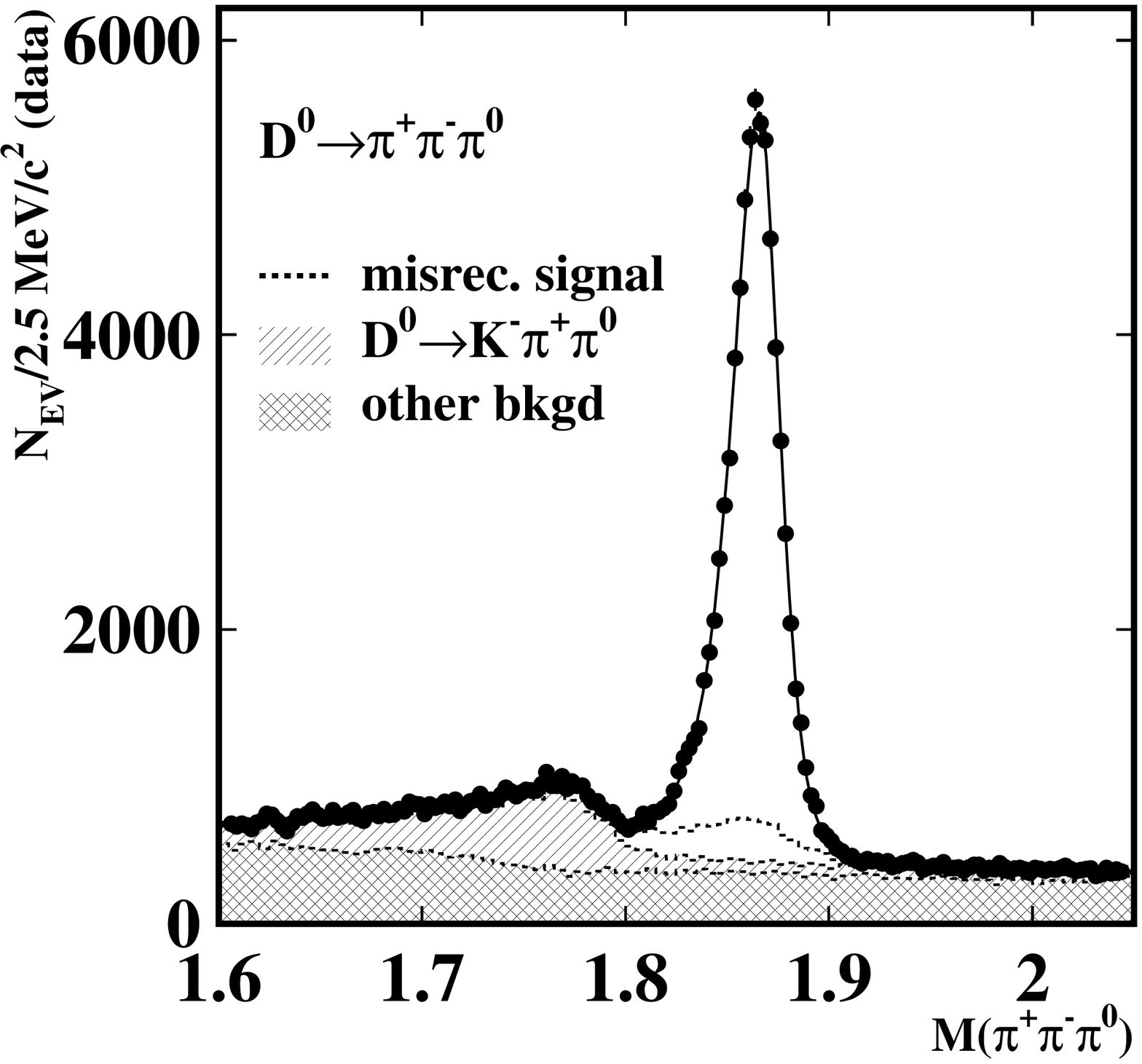,width=0.48\textwidth} 
 \epsfig{figure=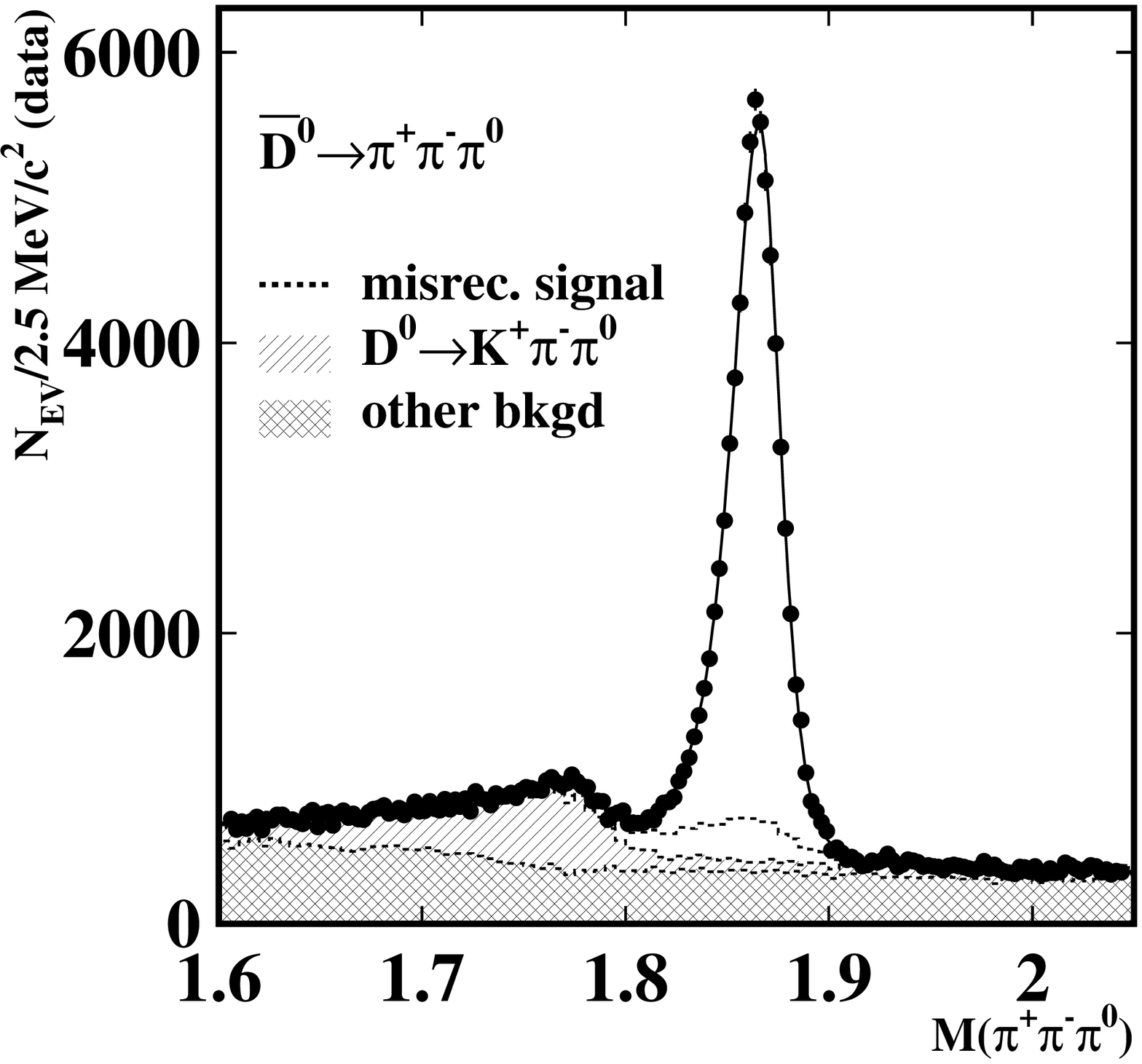,width=0.48\textwidth} 
 \put(-260,180){\bf a)}  
 \put(-30,180){\bf b)} 
 \caption{({\bf a}) $M(\dppp)$ and ({\bf b}) $M(\dbar \to \pi^+\pi^-\pi^0)$ 
 data. Background: misreconstructed signal (dashed line), 
 $\dkpp$ with misidentified kaon (shaded histogram) and other
 sources (hatched histogram). Events from the $M(D^0)$ signal region 
 (1.79 to 1.91 GeV/$c^2$) are selected.}
\end{figure}

A detector bias may exist that leads to different efficiencies for reconstructing 
positively and negatively charged tracks. This may be due to charge-dependent 
effects such as opposite signs of the Lorentz angle with respect to the 
curvature of tracks in the CDC, and a difference in nuclear 
interactions with the detector material for positive and negative tracks. 
The former is partially taken into account by generating signal MC samples for 
$D^{*+} \to D^0 \pi^+_{\rm tag}$ and $D^{*-} \to \dbar \pi^-_{\rm tag}$ 
separately. However, nuclear interactions of charged tracks with the detector 
material are imperfectly simulated. This fact also causes a systematic difference 
between tracking efficiencies for positive and negative particles and has to be 
taken into account. \\ 

Since $D^0$ and $\dbar$ are distinguished only by $\pi^{\pm}_{\rm tag}$, 
and the neutral $D$ meson decays are charge and particle-type balanced, the 
uncertainties of the reconstruction efficiencies of the pions originating from $D^0$ 
do not affect the result. We consider the uncertainties in the tracking and PID 
efficiencies of the tagging pions as the main source of systematic errors for $A_{CP}$. \\ 

The uncertainty of the tracking efficiency was obtained using the same method used for 
the systematics of the $\dppp$/$\dkpp$ ratio, but in this case, positive and negative 
$\pi_{\rm tag}$'s were treated separately. The calculation of the systematic error takes 
into account the momentum dependence. The errors for $\pi^+_{\rm tag}$ and 
$\pi^-_{\rm tag}$ were propagated to $A_{CP}$ assuming them to be uncorrelated. 
The charge-dependent data/MC PID corrections for $\pi_{\rm tag}$ were obtained using 
independent $D^* \to D^0(K_S \pi^0) \pi_{\rm tag}$ data and MC samples. \\ 

In general, the $D$-meson distribution is an asymmetric function of $\cos(\theta)$ 
(where $\theta$ is the polar angle) due to the interference of virtual $\gamma$ and $Z^0$ 
in the process of $c$-quark pair production. If the detector acceptance in the 
center-of-mass frame were perfectly symmetric, the $\cos(\theta)$ dependent asymmetry of 
$D^0$ and $\dbar$ ($D^+$ and $D^-$ etc.) production would cancel out in the integral over 
$\cos(\theta)$ in a symmetric interval. However, the detector acceptance is not symmetric 
and a possible forward--backward asymmetry ($A_{fb}$) should be taken into account. 
A data sample of $D^0\to K^+K^-,\pi^+\pi^-$ decay events was used to calculate 
$A_{fb}(\cos(\theta))$. 
This function was then used to weight the MC $\dppp$ distribution, 
which was then normalized to the total number of MC $\dppp$ events. 
The calculated value equals 0.15\% and is treated as the systematic uncertainty 
related to the forward--backward asymmetry. 
Other individual sources of systematic uncertainties are listed in Table \ref{t4}. 
Systematic errors for each $D^0$ flavor are calculated similarly to those for 
$\bratio$, propagated to $A_{CP}$, and then added in quadrature. \\ 

\begin{table}[ht!]
\caption{Systematic uncertainties for $A_{CP}$:}  
\vspace{0.5cm}
\label{t4}
\begin{tabular}{|c|c|c|c|c|c|c|c|c|} 
\hline  
Source       & MC stat. & Tracking & Fit   & $K_S$ veto & PID  & Binning & $A_{fb}$ & Total \\ \hline  
$\sigma$, \% & 0.24     & 1.01     & 0.58  & 0.23       & 0.15 & 0.05   & 0.15 & 1.23 \\ \hline 
\end{tabular} \\   
\end{table}


The resulting value of the asymmetry is 
\begin{equation} 
\begin{split} 
& A_{CP} \ = \ (S_{D^0}-S_{\dbar})/(S_{D^0}+S_{\dbar}) \ = \ \\ 
& \ = \ (0.43 \pm 0.41({\rm stat}) \pm 1.01({\rm track})  
 \pm 0.70({\rm other \ syst}))\%  \ = \ \ (0.43 \ \pm \ 1.30)\%. 
\end{split} 
\end{equation} 
This result is consistent with $CP$ conservation in this decay mode; 
its sensitivity is a significant improvement over that of the previous 
measurement, $(1 ^{+10}_{-9})\%$~\cite{cleo2}. 

\section{Summary} 

Using 532 fb$^{-1}$ of data collected with the Belle detector, 
a high-precision measurement of the relative branching fraction
$\bratio \ = \ 0.1012 \pm 0.0004 \pm 0.0018$ has been performed. 
The method applied minimizes possible systematic uncertainties due to the $D^0$ 
decay model. The mode $\dkpp$ is chosen for normalization to avoid most of the 
tracking and particle identification uncertainties. 
We also calculate the value of the time-integrated $CP$ asymmetry to be 
$A_{CP}(\dppp) \ = \ (0.43 \ \pm \ 1.30)\%$, which is consistent 
with zero. The sensitivity is significantly better than 
that of the previous measurement~\cite{cleo2}. \\ 

\section{Acknowledgements}

We thank the KEKB group for the excellent operation of the
accelerator, the KEK cryogenics group for the efficient
operation of the solenoid, and the KEK computer group and
the National Institute of Informatics for valuable computing
and Super-SINET network support. We acknowledge support from
the Ministry of Education, Culture, Sports, Science, and
Technology of Japan and the Japan Society for the Promotion
of Science; the Australian Research Council and the
Australian Department of Education, Science and Training;
the National Natural Science Foundation of China under
contract No.~10575109 and 10775142; the Department of
Science and Technology of India; 
the BK21 program of the Ministry of Education of Korea, 
the CHEP SRC program and Basic Research program 
(grant No.~R01-2005-000-10089-0) of the Korea Science and
Engineering Foundation, and the Pure Basic Research Group 
program of the Korea Research Foundation; 
the Polish State Committee for Scientific Research; 
the Ministry of Education and Science of the Russian
Federation and the Russian Federal Agency for Atomic Energy;
the Slovenian Research Agency;  the Swiss
National Science Foundation; the National Science Council
and the Ministry of Education of Taiwan; and the U.S.\
Department of Energy. 



\begin{thebibliography}{99}
\bibitem{kplus} Unless specified otherwise, both flavors of $D^0$ mesons 
 are implied: $\dkpp$ and $\dbar \to K^+ \pi^- \pi^0$.
\bibitem{dstar} $D^*$'s originate mainly from continuum. 
 Although we do not apply any topological cuts, the yield of 
 $D^*$'s coming from $B$ mesons is negligible: they are rejected by 
 kinematic cuts, mainly by the stringent p$_{\rm cms}$($D^*$) requirement.
\bibitem{kpi} The backward end of the ACC corresponds to 
 $\theta_{\rm lab} \ \sim \ 2.2$ rad, and as a result, 
 kaon--pion separation is less efficient for $\theta_{\rm lab} \ > \ 2.2$ rad.
\bibitem{hyper} $f \ = \ e^{-(y-\alpha)}
 \cdot \frac{\alpha \cdot |(x-x_0)/\sigma|}{2\sqrt{2\sigma} 
 \cdot y \cdot \sqrt{y-\alpha}}$, where $y \ = \ \sqrt{\alpha^2 + 
 \alpha \cdot ((x-x_0)/\sigma)^2}$. The function tends to a Gaussian 
 for $|x-x_0| \ll \sqrt{\alpha} \cdot \sigma$ and has exponential 
 behaviour for $|x-x_0| \gg \sqrt{\alpha} \cdot \sigma$.
\bibitem{about_newcleo}{Our conclusions do not change if instead of the world 
 average value of $\mathcal{B}(D^0 \to K^-\pi^+)$ we use the value of this branching 
 fraction from the recent high precision measurement of CLEO~\cite{newcleo}.}
\end{thebibliography}
\end{document}